\documentclass[10pt,showkeys,showpacs,aps,preprintnumbers]{revtex4}

\usepackage{graphicx}% Include figure files
\usepackage{amssymb}
\usepackage{bm}
\usepackage[sumlimits,intlimits]{amsmath}
\usepackage{amsfonts,amssymb}
\usepackage{amsmath}
\usepackage{bm}
\usepackage{graphics}
\usepackage{fancyhdr}
\usepackage{rotating}
\usepackage{graphicx}
\usepackage{textcomp}
\usepackage{color}

\begin{document}

\author{L. Toledo Sesma}
\email{ltoledo@fisica.ugto.mx}

\author{J. Socorro}
\email{socorro@fisica.ugto.mx}

\author{O. Loaiza}
\email{oloaiza@fisica.ugto.mx}

\title{Time-dependent toroidal compactification proposals and the Bianchi type I model: classical and quantum solutions}

\affiliation{Departamento de F\'isica, DCI. Universidad de Guanajuato. Campus Le\'on.\\
Loma del Bosque No. 103 Col. Lomas del Campestre C.P 37150 A.Postal
E-143\\ Le\'on, Guanajuato, Mexico.}

\begin{abstract}
\vspace{1cm}
\begin{center}
{\bf Abstract}
\end{center}
In this work we construct an effective four-dimensional model by compactifying a ten-dimensional theory of gravity coupled
with a real scalar dilaton field on a time-dependent torus. This approach is applied to anisotropic cosmological Bianchi type
I model for which we study the classical coupling of the anisotropic scale factors with the two real scalar moduli produced by
the compactification process. Under this approach, we present an isotropization mechanism for the Bianchi I cosmological model
through the analysis of the ratio between the anisotropic parameters and the volume of the Universe which in general keeps
constant or runs into zero for late times. We also find that the presence of extra dimensions in this model can accelerate the
isotropization process depending on the momenta moduli values.  Finally, we present some solutions to the corresponding Wheeler-DeWitt (WDW)
equation in the context of Standard Quantum Cosmology.
\end{abstract}

%\draft
\date{\today}
\maketitle

\section{Introduction}
The 2015 release of Planck data has provided a detailed map of cosmic microwave background (CMB)  temperature and polarization
allowing us to detect deviations from an isotropic early Universe \cite{Ade:2015bva}. The evidence given by these data leads us
 to the possibility  to consider that there is no exact isotropy, since there exist small anisotropy deviations of the CMB radiation
 and apparent large angle anomalies. In that context,  there have been recent attempts to fix constraints on such deviations by using the Bianchi
 anisotropic models \cite{Russell:2013oda}. The basic idea behind these models is to consider the present observational anisotropies and
 anomalies as imprints of an early anisotropic phase on the CMB which in turn can be explained by the use of different Bianchi models.
 In particular, Bianchi I model seems to be related to large angle anomalies  \cite{Russell:2013oda}
 (and references therein).\\

In a different context, attempts to understand
diverse aspects of cosmology, as the presence of stable vacua and
inflationary conditions,  in the framework of  supergravity and
string theory  have been considered in the last years \cite{Damour:1994zq,Horne:1994mi,
Banks:1995dp,Damour:2001sa,Banks:1998vs,Kallosh:2001ai,Kallosh:2001du,
Baumann:2014nda}. One of the most interesting features emerging from
these type of models consists on the study of the consequences of
higher dimensional degrees of freedom on the cosmology derived from
four-dimensional effective theories
\cite{Khoury:2001bz,MartinezPrieto:2004fd}. The usual procedure for that is to consider compactification on generalized manifolds,
on which  internal fluxes have back-reacted, altering the smooth Calabi-Yau geometry  and stabilizing all  moduli \cite{Baumann:2014nda}.\\

The goal of the present work is to consider  in a simple model some of the above two perspectives, i.e., we shall consider the presence of
extra dimensions in a Bianchi I model with the purpose to track down the influence of  moduli fields in its isotropization.
For that we shall consider an alternative procedure concerning the role played by moduli. In particular we shall not
consider the presence of fluxes, as in string theory, in order to obtain a moduli-dependent scalar potential in the effective
theory. Rather, we are going to promote some of the moduli to  time-dependent fields by considering  the particular case of a
ten-dimensional gravity coupled to a time-dependent dilaton compactified on a 6-dimensional torus with a time-dependent
K\"ahler modulus. With the purpose  to study the influence of such  fields, we are going to ignore the dynamics of the
complex structure modulos (for instance, by assuming that it is already stabilized by the presence  of a string field in higher
scales). This shall allows us to construct classical effective models with two moduli. \\

However, we are also interested in possible quantum aspects of our model. Quantum implications on cosmology  from more fundamental theories
are expected due to different observations. For instance,  it has been pointed out that the presence of extra dimensions leads to an interesting
connection to the
ekpyrotic model \cite{Khoury:2001wf}, which  generated considerable activity \cite{Kallosh:2001du,Kallosh:2001ai,
Enqvist:2001zk}. The essential ingredient in these models (see for instance \cite{Khoury:2001bz}) is to  consider an effective
action with a graviton and a massless scalar field, the dilaton, describing the evolution of the Universe, while incorporating
some of the ideas of pre-big-bang proposal \cite{Veneziano:1991ek,Gasperini:1992em} in that the evolution of the Universe began
in the far past. On the other hand, it is well known that relativistic theories of gravity such as general relativity or string theories are
invariant under reparametrization of time. Quantization of such theories presents a number of problems of principle known
as ``the problem of time'' \cite{Isham:1992ms,Kuchar:1991qf}. This problem is present in all systems whose classical version is
invariant under time reparametrization, leading to  its absence at the quantum level. Therefore, the formal question involves
how to handle the classical Hamiltonian constraint, $\mathcal{H} \approx 0$, in the quantum theory. Also, connected with the
problem of time is the ``Hilbert space problem'' \cite{Isham:1992ms,Kuchar:1991qf} referring to the not at all obvious
selection of the  inner product of states in quantum gravity, and whether there is a need for such a structure at all.
The above features, as it is well known, point out the necessity to construct a consistent theory of gravity at quantum level.\\
%One promising attempt is string theory where the structure of the internal space plays a relevant role in the construction of
%interesting effective models containing cosmological features as the presence of positive valued minima of the scalar potential
%and  slow-roll inflationary conditions.

Analysis of effective four-dimensional cosmologies derived from
M-theory and Type IIA string theory were considered in
\cite{Billyard:1999dg, Billyard:1999ct, Billyard:1999wu} where
string fluxes are related to the dynamical behaviour of the
solutions.
 A quantum description of the model was studied in    \cite{Cavaglia:2000rx, Cavaglia:2000uu} where a flat Friedmann-Robertson-Walker geometry
  was considered for the extended four-dimensional space-time while a geometry given by $S^1\times T^6$ was assumed for the
  internal seven-dimensional space,  however the dynamical field are in the quintom scheme,
since  one of the fields have a negative energy. Under a similar
perspective, we study the Hilbert space of quantum states on a
Bianchi I geometry with
  two  time-dependent scalar moduli derived from a ten-dimensional effective action containing the dilaton and the K\"ahler parameter from a
  six-dimensional torus compactification. We find that in this case, the wave function of this Universe is represented by two factors, one depending on moduli
  while the second one depending on gravitational fields. This behaviour is a property of all cosmological Bianchi Class A models.\\

The work is organized in the following form. In section II we present the construction of our effective action by
compactification on a time-dependent torus,  while in  Section III we study its Lagrangian and Hamiltonian descriptions using
as toy model the Bianchi type I cosmological model. Section IV is devoted on finding the corresponding classical solutions for
few different cases involving the presence or absence of matter content and the cosmological constant term. Using the classical
solutions found in previous sections, we present an isotropization mechanism for the Bianchi I cosmological model, through the
analysis of the ratio between the anisotropic parameters and the volume of the Universe, showing that in all the cases we have
studied, its value keeps constant or runs into zero for late times. In Section V we present some solutions to the corresponding
Wheeler-DeWitt (WDW) equation in the context of Standard Quantum Cosmology and finally our conclusions are presented in Section
VI.

\section{Effective model}
We start from a ten-dimensional action coupled with a dilaton (which
is the bosonic component common to all superstring theories), which
after dimensional reduction can be interpreted as a Brans-Dicke like
theory \cite{Brans:1961sx}. In the string frame, the effective
action depends on two spacetime-dependent scalar fields: the dilaton
$\Phi(x^{\mu})$ and the K\"ahler modulus $\sigma(x^{\mu})$. For simplicity,  in this work we shall
assume that these fields only depend on time.
The  high-dimensional (effective) theory is therefore given by
\begin{equation}
S=\frac{1}{2 \kappa^2_{10}} \int d^{10}X
\sqrt{-\hat{G}}\,e^{-2\Phi}\,\Bigg[\hat{\mathcal{R}}^{(10)}
+ 4\,\hat{G}^{MN}\nabla_{M}\,\Phi\,\nabla_{N}\,\Phi\Bigg] + \int
d^{10}X\,\sqrt{-\hat{G}}\,\mathcal{\hat L}_{m}
\label{action}
\end{equation}
where
all quantities $\hat q$ refer  to the string frame while
the ten-dimensional metric is described by
\begin{equation}
ds^2=\hat{G}_{MN}\,dX^{M}\,dX^{N}= \hat{g}_{\mu\nu}\,dx^{\mu}dx^{\nu} + h_{mn}\,dy^{m}dy^{n},
\label{allgemeinemetric}
\end{equation}
where $M, N, P, \ldots$ are the indices of the ten-dimensional space, greek indices $\mu, \nu,\ldots =0,\ldots, 3$ and latin
indices $m, n, p,\ldots=4,\ldots,9$ correspond to the external and internal space, respectively. We will assume that the
six-dimensional internal space has the form of a torus with a metric  given by
\begin{equation}
h_{mn}=e^{-2\sigma(t)}\,\delta_{mn},
\label{intmet}
\end{equation}
with $\sigma$ a real parameter.\\

Dimensionally reducing the first term in \eqref{action} to four-dimensions in the Einstein frame (see Appendix A for details), gives
%We are interested to describe a four-dimensional model; by realizing
%this task we need to reduce the ten-dimensional equations of motion
%to four dimensions and in the Einstein frame. In this way, the
%effective dilaton $\phi$ is also a spacetime -dependent field.
%Following this notion, we have that the action  in
%four dimensions reads
%
\begin{equation}
S_4=\frac{1}{2\kappa^2_{4}}\int d^4x \sqrt{-g}\,\biggl(\mathcal{R}  -2\,g^{\mu\nu}\,\nabla_{\mu}\phi\nabla_{\nu}\phi
- 96\,g^{\mu\nu}\,\nabla_{\mu}\sigma\nabla_{\nu}\sigma - 36\,g^{\mu\nu}\,\nabla_{\mu}\phi\nabla_{\nu}\sigma\biggr),
\label{endeaction}
\end{equation}
where $\phi=\Phi+\frac{1}{2}ln(\hat{V})$, with $\hat{V}$ given by
\begin{equation}
\hat{V}=e^{6\sigma(t)}Vol(X_6)=\int d^6y.
\end{equation}
By considering only a time-dependence on the moduli, one can notice that for the internal volume $Vol(X_6)$ to be small, the modulus $\sigma(t)$
should be a monotonic increasing function on time (recall that $\sigma$ is a real parameter), while $\hat{V}$ is time and moduli independent.\\

Now, concerning the second term in \eqref{action}, we shall requiere to properly define the ten-dimensional stress-energy tensor $\hat{T}_{MN}$.
In the string frame it takes the form
\begin{equation}
\hat{T}_{MN}=
 \begin{pmatrix}
  \hat{T}_{\mu\nu} & 0  \\
   0 & \hat{T}_{mn} \\
 \end{pmatrix},
\label{matrixform}
\end{equation}
where $\hat{T}_{\mu\nu}$ and  $\hat{T}_{mn}$ denote the four and six-dimensional components of $\hat{T}_{MN}$.  Observe that we are not
 considering mixing components of $\hat{T}_{MN}$ among internal and external components although non-constrained expressions for $\hat{T}_{MN}$
 have been considered in \cite{Wesley:2008fg}. In the Einstein frame the four-dimensional components are given by
\begin{equation}
T_{\mu\nu}= e^{2\phi}\,\hat{T}_{\mu\nu}.
\label{hauptarroz}
\end{equation}
It is important to remark that there are some dilemmas about what the best frame to describe the
gravitational theory is. Here in this work we have taken the
Einstein frame.  Useful references to
find a discussion about String and Einstein frames and its
relationship in the cosmological context are
\cite{Casas:1991zf,Copeland:1994vi,Lidsey:1999mc,gasperini2007elements,
Rasouli:2014dxa}. \\

Now with the expression \eqref{endeaction} we
proceed to build the Lagrangian and the Hamiltonian of the theory at
the classical regime employing the anisotropic cosmological Bianchi
type I model. The moduli fields will satisfy the Klein-Gordon like
equation in the Einstein frame  as an
effective theory.

\section{Classical Hamiltonian}
In order to construct the classical Hamiltonian,
we are going to assume that the background of
the extended space is described by a cosmological Bianchi type
I model. For that, let us recall here the canonical
formulation in the ADM formalism of the diagonal Bianchi Class A
cosmological models; the metric has the form
\begin{equation}
ds^2= -N(t)dt^2 + e^{2\Omega(t)}\, (e^{2\beta(t)})_{ij}\,\omega^i \, \omega^j,
\label{met}
\end{equation}
where $N(t)$ is the lapse function, $\beta_{ij}(t)$ is a $3\times3$ diagonal matrix, $\beta_{ij}=\text{diag}(\beta_++ \sqrt{3}
\beta_-,\beta_+- \sqrt{3} \beta_-, -2\beta_+)$, $\Omega(t)$ and $\beta_\pm$ are scalar functions, known as Misner variables,
$\omega^i$ are one-forms that characterize each cosmological Bianchi type model \cite{Ryan:1975jw}, and obey the form
$d\omega^i= \frac{1}{2}C^i_{jk} \omega^j \wedge \omega^k$, with $C^i_{jk}$ the structure constants of the corresponding model.
The one-forms for the Bianchi type I model are $\omega^1=dx, \omega^2=dy$, and $\omega^3=dz$. So, the corresponding metric of
the Bianchi type I in Misner's parametrization has the form
\begin{equation}
ds^2_I=-N^{2}\,dt^2 +R_{1}^2\,dx^2 + R_{2}^2\,dy^2 + R_{3}^2\,dz^2,
\label{metbia}
\end{equation}
where $\{R_{i}\}_{i=1}^3$ are the anisotropic radii and they are given by
\begin{equation*}
R_1=e^{\Omega +\beta_+ +\sqrt{3}\beta_-}, \qquad R_2=e^{\Omega +\beta_+
-\sqrt{3}\beta_-}, \qquad  R_3=e^{\Omega -2\beta_+ }, \qquad V=R_{1}R_{2}R_{3}=e^{3\Omega},
\end{equation*}
where $V$ is the volume function of this model.\\

The Lagrangian density, with a matter content given by
 a barotropic perfect fluid and a cosmological
term, has a structure , corresponding to an
energy-momentum tensor of perfect fluid \cite{Socorro:2003fn,Chowdhury:2006pk,Nojiri:1998wb,Nojiri:1999iv}
\begin{equation*}
T_{\mu\nu}=(p+\rho) u_\mu u_\nu + p g_{\mu \nu}
\end{equation*}
that satisfies the conservation law $\nabla_\nu T^{\mu\nu}=0$.  Taking the equation of state $p=\gamma \rho$ between the energy
density and the pressure of the comovil fluid, a solution is given by  $\rho=C_\gamma e^{-3(1+\gamma) \Omega}$ with  $C_\gamma$ the
corresponding constant for different values of $\gamma$ related to the Universe evolution stage. Then, the Lagrangian density reads
\begin{equation*}
\mathcal{L}_{mat}=16 \pi G_N \sqrt{-g} \rho+2\sqrt{-g}
\Lambda=16N\pi G_N C_\gamma e^{-3(1+\gamma) \Omega} +2N\Lambda,
e^{3\Omega}
\end{equation*}
while the Lagrangian that describes the fields dynamics is given by
\begin{equation}
\mathcal{L}_I= \frac{e^{3\Omega}}{N}\left[ 6\,\dot{\Omega}^2 -
6\,\dot{\beta}^2_{+} - 6\,\dot{\beta}^2_{-} + 96\,\dot{\sigma}^2 +
36\,\dot{\phi}\,\dot{\sigma} + 2\,\dot{\phi}^2 +
16\,\pi\,G\,N^2\,\rho + 2\,\Lambda\,N^2\right], \label{lagra-i}
\end{equation}
using the standard definition of the momenta,
$\Pi_{q^\mu}=\frac{\partial {\cal L}}{\partial \dot q^\mu}$, where
$\rm q^\mu = (\Omega, \beta_+, \beta_-,\phi,\sigma)$, we obtain
\begin{eqnarray}
\rm \Pi_\Omega &=&\rm \frac{12 e^{3\Omega} \dot \Omega}{N},
\qquad\qquad\qquad \dot
\Omega=\frac{N e^{-3\Omega}\Pi_\Omega}{12}, \nonumber\\
\rm \Pi_\pm &=&\rm - \frac{12 e^{3\Omega} \dot \beta_\pm}{N},
\qquad\quad \quad\dot
\beta_\pm = -\frac{N e^{-3\Omega}\Pi_{\beta_\pm}}{12}, \nonumber\\
\rm \Pi_\phi &=& \rm \frac{e^{3\Omega}}{N}\left[ 36 \dot \sigma + 4
\dot \phi \right], \qquad \dot \phi=\frac{N
e^{-3\Omega}}{44}\left[3\Pi_\sigma -16 \Pi_\phi \right], \nonumber\\
\rm \Pi_\sigma &=& \rm \frac{e^{3\Omega}}{N}\left[ 192 \dot \sigma +
36 \dot \phi \right], \qquad \dot \sigma=\frac{N
e^{-3\Omega}}{132}\left[9\Pi_\phi - \Pi_\sigma \right], \nonumber
\end{eqnarray}
and, introducing them into the Lagrangian density, we obtain the
canonical Lagrangian as ${\cal L}_{canonical} =\Pi_{q^\mu} \dot
q^\mu - N{\cal H}$. When we perform the variation of this canonical
lagrangian with respect to N, $\frac{\delta {\cal
L}_{canonical}}{\delta N} =0$, implying the constraint ${\cal
H}_I=0$. In our model the only constraint corresponds to Hamiltonian
density, which is weakly zero.  So, we obtain the Hamiltonian
density for this model
\begin{equation}
\mathcal{H}_I=\frac{e^{-3\Omega}}{24}\left[\Pi_\Omega^2 - \Pi_+^2 -
\Pi_-^2 - \frac{48}{11}\Pi_\phi^2 +
\frac{18}{11}\Pi_\phi\,\Pi_\sigma - \frac{1}{11}\Pi_\sigma^2 -
384\,\pi\,G_{N}\,\rho_{\gamma}\,e^{3(1-\gamma)\Omega} -
48\,\Lambda\,e^{6\Omega}\right]. \label{ham}
\end{equation}
We introduce a new set of variables in the gravitational part given by
$e^{\beta_1+\beta_2+\beta_3}=e^{3\Omega}=V$ which corresponds to the
volume of the Bianchi type I Universe, in similar way that the flat
Friedmann-Robetson-Walker metric (FRW) with a scale factor.
This new set of variables  depends on Misner variables as
\begin{eqnarray}
\beta_1&=& \Omega+\beta_+ +\sqrt{3} \beta_-,\nonumber\\
\label{newvaror}\beta_2&=& \Omega+\beta_+ -\sqrt{3} \beta_-,\\
\beta_3&=& \Omega-2\beta_+,\nonumber
\end{eqnarray}
from which the Lagrangian density \eqref{lagra-i} can be transformed as
\begin{equation}
\mathcal{L}_I=\frac{e^{\beta_1+\beta_2+\beta_3}}{N}\biggl(2\,\dot{\phi}^2
+ 36\,\dot{\phi}\,\dot{\sigma} + 96\,\dot{\sigma}^2 +
2\,\dot{\beta}_{1}\,\dot{\beta}_{2} +
2\,\dot{\beta}_{1}\,\dot{\beta}_{3} +
2\,\dot{\beta}_{3}\,\dot{\beta}_{2} + 2\,\Lambda\,N^2 + 16\pi
GN^2\rho_{\gamma}\,e^{-(1 + \gamma)(\beta_{1} + \beta_{2} +
\beta_{3})}\biggr),
\end{equation}
and the Hamiltonian density reads
\begin{align}
\mathcal{H}_{I}=\frac{1}{8}\,e^{-(\beta_{1} + \beta_{2} +
\beta_{3})}\biggl[-\Pi^2_{1} - \Pi^2_{2} &- \Pi^2_{3} +
2\,\Pi_{1}\Pi_{2} + 2\,\Pi_{1}\Pi_{3} + 2\,\Pi_{2}\Pi_{3} -
\frac{16}{11}\,\Pi^2_{\phi}
+ \frac{6}{11}\,\Pi_{\phi}\,\Pi_{\sigma} - \frac{1}{33}\,\Pi^2_{\sigma}\nonumber\\
&- 128\,\pi\,G\,\rho_{\gamma}\,e^{(1 - \gamma)(\beta_{1} + \beta_{2} + \beta_{3})}
- 16\,\Lambda\,e^{2(\beta_{1} + \beta_{2} + \beta_{3})}\biggr].
\label{hami}
\end{align}
So far, we have built the classical Hamiltonian density from a
higher dimensional theory; this Hamiltonian contains a barotropic
perfect fluid, we have explicitly added. The next step is to analyze
three different cases involving terms in the classical Hamiltonian
\eqref{hami},
and find solutions to each of them in the classical regime.\\

%%%%%%%%%%%%%%%%%%%%%%%
\subsection{Isotropization}
%%%%%%%%%%%%%%%%%%%%%%%
 The current observations of the cosmic background radiation set a very
stringent limit to the anisotropy of the Universe \cite{Martinez:1995aa}, therefore it is important to consider the
anisotropy of the solutions.  We shall denote through all our study derivatives of functions $F$ with respect to $\tau=Nt$ by $F'$
(the following analysis is similar to this presented recently in the K-essence theory, since the corresponding action is
similar to this approach \cite{Socorro:2014ama}). Recalling the Friedmann like equation to the Hamiltonian density
\begin{equation}
\Omega^{\prime 2}=\beta_+^{\prime 2}+\beta_-^{\prime 2} +
16\,\sigma^{\prime 2} + \frac{1}{3}\,\phi^{\prime 2} +6\,\phi^\prime
\sigma^\prime + \frac{8}{3}\pi G \mu_\gamma
e^{-3(1+\gamma)\Omega}+\frac{\Lambda}{3}, \label{ome}
\end{equation}
we can see that isotropization is achieved when the terms with $\beta_{\pm}^{\prime2}$ go to zero or are negligible with
respect to the other terms in the differential equation. We find in the literature the criteria for isotropization, among
others, $(\beta_+^{\prime2}+\beta_-^{\prime 2})/H^2\quad \rightarrow \quad 0$,
$(\beta_+^{\prime 2}+\beta_-^{\prime 2})/\rho \quad \rightarrow \quad  0$ , that are consistent with our above remark. In the
present case the comparison with the density should include the contribution of the scalar field.  We define an anisotropic
density $\rho_a$, that is proportional to the shear scalar,
\begin{equation}
\rho_a = \beta_+^{\prime 2}+\beta_-^{\prime 2},
\end{equation}
and will compare it with $\rho_\gamma$, $\rho_\phi$, and $\Omega^{\prime 2}$. From the Hamilton analysis we now see that
\begin{eqnarray}
\dot \Pi_\phi&=&0, \qquad \rightarrow \qquad
\Pi_\phi=p_\phi=constant,\nonumber\\
\dot \Pi_\sigma&=&0, \qquad \rightarrow \qquad
\Pi_\sigma=p_\sigma=constant,\nonumber\\
\dot \Pi_+&=&0, \qquad \rightarrow \qquad
\Pi_+=p_+=constant,\nonumber\\
\dot \Pi_-&=&0, \qquad \rightarrow \qquad
\Pi_-=p_-=constant,\nonumber
\end{eqnarray}
and the kinetic energy for the scalars fields $\phi$ and $\sigma$
are proportional to $e^{-6\Omega}$. This can be seen from definitions
of  momenta associated to the Lagrangian \eqref{lagra-i} and
the above equations. Then,  defining $\kappa_\Omega$ as  $\rm \kappa_\Omega^2\sim p_+^2 + p_-^2 +
\frac{16}{132^2}\left(p_\sigma-9p_\phi\right)^2+\frac{1}{3\cdot44^2}\left(3p_\sigma
-16p_\phi \right)^2+\frac{1}{22\cdot44} \left(p_\sigma - 9p_\phi
\right)\left(3p_\sigma - 16p_\phi\right)$, we have that
\begin{equation}
\rho_a \sim  e^{-6\Omega}, \qquad \rho_{\phi,\sigma} \sim
e^{-6\Omega},\qquad \Omega^{\prime 2} \sim
\frac{\Lambda}{3}+{\kappa_\Omega
}^2e^{-6\Omega}+b_{\gamma}e^{-3(1+\gamma)\Omega}, \quad
b_\gamma=\frac{8}{3}\pi G_N \mu_\gamma.
\end{equation}
With the use of these parameters, we find that the following ratios :
\begin{equation}
\frac{\rho_a }{\rho_{\phi,\sigma}}\sim  constant,\qquad \frac{\rho_a
}{\rho_\gamma}\sim      e^{3\Omega(\gamma-1)},\qquad \frac{\rho_a
}{\Omega^{\prime 2      }}\sim \frac{1}{{\kappa_\Omega }^2  +
\frac{\Lambda}{3} e^{6\Omega}+b_{\gamma}e^{3(1-\gamma)\Omega}}.
\end{equation}
We observe that for an expanding Universe the anisotropic density is dominated by the fluid density (with the exception of the
stiff fluid) or  by the $\Omega^{\prime 2}$ term and then \emph{at late times the isotropization is obtained since the above ratios tends to zero}.

\section{Case of interest}
In this section we present the classical solutions for the Hamiltonian density of the Lagrangian \eqref{lagra-i} we have previously
built in terms of a new set of variables
\eqref{newvaror}, focusing on three different cases. We start our analysis
on the vacuum case, and on the case with a cosmological term $\Lambda$. Finally we shall analyze the general case considering
matter content and a cosmological term.\\

\subsection{Vacuum case}
To analyze the vacuum case, we will take in the Hamiltonian density \eqref{hami} that $\rho_\gamma=0$ and $\Lambda=0$,
obtaining that
\begin{equation}
\mathcal{H}_{I_{vac}} = \frac{1}{8}\,e^{-(\beta_{1} + \beta_{2} +
\beta_{3})}\biggl[-\Pi^2_{1} - \Pi^2_{2} - \Pi^2_{3} +
2\,\Pi_{1}\Pi_{2} + 2\,\Pi_{1}\Pi_{3} + 2\,\Pi_{2}\Pi_{3} -
\frac{16}{11}\,\Pi^2_{\phi} + \frac{6}{11}\,\Pi_{\phi}\,\Pi_{\sigma}
- \frac{1}{33}\,\Pi^2_{\sigma}\biggr]. \label{hami-sin}
\end{equation}
%\begin{equation}
%\mathcal{H}_{I_{vac}} = \frac{1}{8}\,e^{-(\beta_{1} + \beta_{2} +
%\beta_{3})}\biggl[-\Pi^2_{1} - \Pi^2_{2} - \Pi^2_{3} +
%2\,\Pi_{1}\Pi_{2} + 2\,\Pi_{1}\Pi_{3} + 2\,\Pi_{2}\Pi_{3} -
%b_0\biggr]. \label{hami-sin}
%\end{equation}

The Hamilton equations for the coordinates fields $\rm \dot
q_i=\frac{\partial {\cal H}}{\partial P_i}$ and the corresponding
momenta $\rm \dot P_i=\frac{\partial {\cal H}}{\partial q_i}$
becomes
\begin{eqnarray}
\rm \dot \Pi_i&=&\rm -{\cal H}\equiv 0, \qquad \Rightarrow \qquad \Pi_i=constant, \nonumber\\
\rm \dot \Pi_\phi &=& 0, \qquad \Rightarrow \qquad
\Pi_\phi=constant,
\nonumber\\
\rm \dot \Pi_\sigma &=& 0, \qquad \Rightarrow \qquad
\Pi_\sigma=constant, \nonumber\\
\beta_{1}'& = &\frac{1}{4}\,e^{-(\beta_1+\beta_2+\beta_3)}\,\left[-\Pi_1 +\Pi_2+\Pi_3 \right], \nonumber\\
\beta_{2}'& = &\frac{1}{4}\,e^{-(\beta_1+\beta_2+\beta_3)}\,\left[-\Pi_2 +\Pi_1+\Pi_3 \right], \\
\beta_{3}'& = &\frac{1}{4}\,e^{-(\beta_1+\beta_2+\beta_3)}\,\left[-\Pi_3 +\Pi_1+\Pi_2 \right], \nonumber\\
\phi'& = &\phi_{0}\,e^{-(\beta_1+\beta_2+\beta_3)}, \nonumber\\
\sigma'& = &\sigma_{0}\,e^{-(\beta_1+\beta_2+\beta_3)}.\nonumber
 \end{eqnarray}
The gravitational momenta are constant by mean the Hamiltonian
constraint, first line in the last equation. In this way, the
solution for the sum of $\beta_i$ functions become
\begin{equation*}
\beta_1+\beta_2+\beta_3=\ln\left[\epsilon \tau
+b_0\right],\qquad \epsilon= b_1+b_2+b_3,
\end{equation*}
with $b_i=\Pi_i$ and $b_0$ an integration constant.
%In order to obtain the last expression we have taken the following $\prime=\frac{d}{d\tau}=\frac{d}{Ndt}$ to denote time derivative.
Therefore, Misner variables are expressed as
\begin{equation}
\Omega=\ln\left(\epsilon\tau + b_0\right)^{\frac{1}{3}}, \qquad
\beta_{+}=\ln\left(\epsilon\tau + b_0\right)^{\frac{\epsilon - 3\,b_{3}}{6\,\epsilon}}, \qquad
\beta_{-}=\ln\left(\epsilon\tau + b_0\right)^{\frac{b_{3} + 2\,b_{1} - \epsilon}{2\,\sqrt{3}\,\epsilon}},
\label{tolatetime}
\end{equation}
while moduli fields are given by
\begin{equation*}
\phi=\frac{\phi_{0}}{\epsilon}\,\ln\left(\epsilon\tau + b_{0}\right), \qquad
\sigma=\frac{\sigma_{0}}{\epsilon}\,\ln\left(\epsilon\tau + b_{0}\right),
\end{equation*}
where $\phi_0=\Pi_\phi$ and $\sigma_0=\Pi_\sigma$. Notice that in
this case, the associated external volume given by $e^{3\Omega}$ and
the moduli fields $\phi$ and $\sigma$ are all logarithmically
increasing functions on time, implying for the later that the
internal volume shrinks in size for lately times, as expected, while
the four-dimensional Universe expands into an isotropic flat
space-time. The parameter $\rho_a$, goes to zero at very late times
independently of $b_0$.

%//--------------------------------------------------------------------------------------------------------------

\subsection{Cosmological term $\Lambda$}\label{CTL}
The corresponding Hamiltonian density becomes
\begin{align}
\mathcal{H}_{I} = \frac{1}{8}\,e^{-(\beta_1+\beta_2+\beta_3)}
\biggl[-\Pi^2_{1} - \Pi^2_{2} - \Pi^2_{3} + 2\,\Pi_{1}\Pi_{2} +
2\,\Pi_{1}\Pi_{3} + 2\,\Pi_{2}\Pi_{3} - \frac{16}{11}\,\Pi^2_{\phi}
+ \frac{6}{11}\,\Pi_{\phi}\,\Pi_{\sigma}
\nonumber\\
- \frac{1}{33}\,\Pi^2_{\sigma} -
16\,\Lambda\,e^{2(\beta_1+\beta_2+\beta_3)}\bigg],
\label{hami-lambda}
\end{align}
and the corresponding Hamilton's equations are
\begin{eqnarray}
\Pi_{1}'=\Pi_{2}'=\Pi_{3}'& = & 4\,\Lambda\,e^{\beta_1+\beta_2+\beta_3}, \label{p1i-l}\nonumber\\
\beta_{1}'& = &\frac{1}{4}\,e^{-(\beta_1+\beta_2+\beta_3)}\,\left[-\Pi_1 +\Pi_2+\Pi_3 \right], \label{beta1p-l}\nonumber\\
\beta_{2}'& = &\frac{1}{4}\,e^{-(\beta_1+\beta_2+\beta_3)}\,\left[-\Pi_2 +\Pi_1+\Pi_3 \right], \label{beta2p-l}\\
\beta_{3}'& = &\frac{1}{4}\,e^{-(\beta_1+\beta_2+\beta_3)}\,\left[-\Pi_3 +\Pi_1+\Pi_2 \right], \label{beta3p-l}\nonumber\\
\phi'& = &\phi_{0}\,e^{-(\beta_1+\beta_2+\beta_3)}, \label{phi1i-l}\nonumber\\
\sigma'& = &\sigma_{0}\,e^{-(\beta_1+\beta_2+\beta_3)},\label{sigma1i-l}\nonumber
\end{eqnarray}
where the constants $\phi_0$ and $\sigma_0$ are the same as in previous case. To solve the last system of equations we shall take the following ansatz
\begin{equation}
\Pi_1=\Pi_2+\delta_2=\Pi_3+\delta_3,
\label{momentas-l}
\end{equation}
with $\delta_2$ and $\delta_3$ constants. By substituing  into the Hamiltonian  \eqref{hami-lambda} we find a differential equation for the
momenta $\Pi_1$ given by
\begin{equation}
\frac{1}{\Lambda}\Pi_1^{\prime 2} - 3\,\Pi_1^2 + \nu\,\Pi_1 + \delta_1=0,
\end{equation}
where the corresponding constants are
\begin{equation}
\nu=2\,(\delta_2 +\delta _3), \qquad
\delta_1(q_0)=(\delta_2-\delta_3)^2 - q_0,
\end{equation}
with the constant (related  to the
momenta scalar field)
\begin{equation}
q_0= \frac{16}{11}\,\Pi^2_{\phi} -
\frac{6}{11}\,\Pi_{\phi}\,\Pi_{\sigma}
+\frac{1}{33}\,\Pi^2_{\sigma}.
\label{q0}
\end{equation}
The solution for $\Pi_{1}$ is thus
\begin{equation}
\Pi_1=\frac{\nu}{6} + \frac{1}{6}\,\sqrt{\nu^2 + 12\,\delta_1}\,\cosh\left(\sqrt{3\,\Lambda}\,\tau\right).
\label{pionesolu}
\end{equation}
Hence, using the last result \eqref{pionesolu} in the rest of equations in \eqref{beta1p-l} we obtain that
\begin{equation}
\beta_1 + \beta_2 + \beta_3 = \ln\left(\frac{1}{8}\,\sqrt{\frac{\nu^2 + 12\,\delta_{1}}{3\,\Lambda}}\,
\sinh(\sqrt{3\,\Lambda}\,\tau)\right),
\end{equation}
and the corresponding solutions to the moduli fields are
\begin{eqnarray}
\phi& = &\frac{8\,\phi_2}{\sqrt{\nu^2 + 12\,\delta_{1}}}\,\ln\biggl(\tanh\biggl(\frac{\sqrt{3\,\Lambda}}{2}\,\tau\biggr)\biggr),\\
\sigma& = &\frac{8\,\sigma_2}{\sqrt{\nu^2 + 12\,\delta_{1}}}\,\ln\biggl(\tanh\biggl(\frac{\sqrt{3\,\Lambda}}{2}\,\tau\biggr)\biggr).\nonumber
\end{eqnarray}
With the last results we see that the solutions to the Misner variables $(\Omega,\beta_\pm)$ are given by
\begin{eqnarray}
\Omega& = &\frac{1}{3}\,\ln\left(\frac{1}{8}\,\sqrt{\frac{\nu^2 + 12\,\delta_{1}}{3\,\Lambda}}\sinh(\sqrt{3\,\Lambda}\,\tau)\right)
,\nonumber\\
\beta_{+}& = &\frac{2}{3}\,\frac{(\delta_{2} - 2\,\delta_{3})}{\sqrt{\nu^2 + 12\,\delta_{1}}}\,
\ln\,\tanh\left(\frac{\sqrt{3\,\Lambda}}{2}\,\tau\right),\label{tolatetime-1}\\
\beta_{-}& = &-\frac{2}{\sqrt{3}}\,\frac{\delta_{2}}{\sqrt{\nu^2 +
12\,\delta_{1}}}\,\ln\,\tanh\left(\frac{\sqrt{3\,\Lambda}}{2}\,\tau\right).\nonumber
\end{eqnarray}
and the isotropization parameter $\rho_a$ is given by
\begin{equation}
\rho_a=\frac{\Lambda}{3}\frac{(\delta_2-2\delta_3)^2+\delta_2^2}{\nu^2+12((\delta_2-\delta_3)^2-q_0)}\, \coth^2\left(\frac{\sqrt{3\Lambda}}{2}\tau\right).
\end{equation}
Notice that the external four-dimensional volume expands as time
$\tau$ runs and it is affected wether $q_0$ is positive or negative.
This in turns is determined by how faster evolves the moduli fields
with respect to each other (i.e., wether $\Pi_\phi$ is larger or
smaller than $\Pi_\sigma$). Therefore,  isotropization is reached
independently of the values of $q_0$ but expansion  of the Universe
(and shrinking of the internal volume) is affected by it.

%
%\begin{equation}
%V\sim V_0+ \Delta V(b_0, \tau)=( V_0-\frac{3}{4\sqrt{3\Lambda}}\frac{1}{A_1}b_0) sinh(\sqrt{3\Lambda}\tau),
%\end{equation}
%

%%------------------------------------------------------------------------------------------------------------------------------------

\subsection{Matter content and cosmological term}
For this case, the corresponding Hamiltonian density becomes the
equation \eqref{hami}. So,  using the Hamilton equation, we can see
that the momenta associated to the scalars fields $\phi$ and
$\sigma$ are constants, and we shall label these constants by
$\phi_{1}$ and $\sigma_{1}$, respectively.
\begin{eqnarray}
\Pi_{1}^\prime=\Pi_{2}^\prime=\Pi_{3}^\prime& = &4\,\Lambda\,e^{\beta_1+\beta_2+\beta_3} + 16\,\pi\,G\,(1-\gamma)\,\rho_{\gamma}\,e^{-\gamma(\beta_1+\beta_2 +\beta_3)}, \label{p1i}\nonumber\\
\beta_1^\prime& = &\frac{1}{4}\,e^{-(\beta_1+\beta_2+\beta_3)}\,\left[-\Pi_1 +\Pi_2+\Pi_3 \right], \label{beta1p}\nonumber\\
\beta_2^\prime& = &\frac{1}{4}\,e^{-(\beta_1+\beta_2+\beta_3)}\,\left[-\Pi_2 +\Pi_1+\Pi_3 \right], \label{beta2p}\\
\beta_3^\prime& = &\frac{1}{4}\,e^{-(\beta_1+\beta_2+\beta_3)}\,\left[-\Pi_3 +\Pi_1+\Pi_2 \right], \label{beta3p}\nonumber\\
\phi^\prime& = &\phi_{2}\,e^{-(\beta_1+\beta_2+\beta_3)}, \label{phi1i}\nonumber\\
\sigma^\prime& =
&\sigma_{2}\,e^{-(\beta_1+\beta_2+\beta_3)}.\nonumber
\label{sigma1i}
\end{eqnarray}
Follow the same steps as the section \ref{CTL}, we see that is
possible to relate the momenta associated to $\beta_{1}$,
$\beta_{2}$, and $\beta_{3}$ in the following way
\begin{equation}
\Pi_1=\Pi_2+k_1=\Pi_3+k_2 \label{momentas}
\end{equation}
also, the differential equation for fields $\phi$ and $\sigma$ can
be reduced to quadratures as
\begin{equation}
\phi=\phi_{2}\,\int e^{-(\beta_1+\beta_2+\beta_3)} d\tau,
\qquad \sigma=\sigma_{2}\,\int
e^{-(\beta_1+\beta_2+\beta_3)} d\tau. \label{solutions-fields}
\end{equation}

We proceed to consider specific values for $\gamma$.

\subsubsection{$\gamma=\pm 1$ cases}
Let us introduce
the generic parameter
\begin{equation}
\lambda = \left\{
\begin{tabular}{lc}
$4 \Lambda$,& $\gamma=1$\\
$4 \Lambda+32\pi G \rho_{-1}$,& $\gamma=-1$
\end{tabular}
\right.
\end{equation}
for which $\Pi^\prime_1=\lambda\,e^{\beta_1+\beta_2+\beta_3}$. Substituting this into the Hamiltonian \eqref{hami} we find the
differential equation for the momenta $\Pi_1$ as
\begin{equation}
\frac{4}{\lambda}\,\Pi_1^{\prime 2} - 3\,\Pi_1^2 + \nu\,\Pi_1 + \delta_1=0,
\end{equation}
%where the constants $\kappa$ and $k_{3}$ are given by
%\begin{equation}
%\kappa=2\,(k_1 + k_2), \qquad k_3=(k_1 - k_2)^2 - \phi_0,
%\end{equation}
%and $\phi_0=\frac{16}{11}\,\phi_1^2 - \frac{6}{11}\,\phi_1\,\sigma_1 + \frac{1}{33}\,\sigma_1^2$.
with  the solution given by
$\Pi_1$ is
\begin{equation}
\Pi_1=\frac{\nu}{6} + \frac{1}{6}\,\sqrt{\nu^2 + 12\,\delta_1}\,\cosh\left(\frac{\sqrt{3\,\lambda}}{2}\,\tau\right).
\label{solpiu}
\end{equation}
On the other hand, using the last result \eqref{solpiu} we obtain from Hamilton equations that
\begin{equation}
\beta_1 + \beta_2 + \beta_3=\,\ln\left(\frac{1}{4}\,\sqrt{\frac{\delta_3 + 12\,\delta_1}{3\,\lambda}}\,
\sinh\left(\frac{\sqrt{3\,\lambda}}{2}\,\tau\right)\right).
\end{equation}
The corresponding solutions to the moduli fields \eqref{solutions-fields} can be found as
\begin{equation}
\,\phi = \frac{8\,\phi_{2}}{\sqrt{\nu^2 + 12\,\delta_1}}\,\ln\,\tanh\left(\frac{\sqrt{3\,\lambda}}{4}\,\tau\right),
\qquad
\,\sigma = \frac{8\,\sigma_{2}}{\sqrt{\nu^2 + 12\,\delta_1}}\,\ln\,\tanh\left(\frac{\sqrt{3\,\lambda}}{4}\,\tau\right).
\label{fieldsequatmod}
\end{equation}
 The solutions to the Misner variables \eqref{newvaror}, $(\Omega,\beta_\pm)$ are given
by
\begin{eqnarray}
\Omega& = &\frac{1}{3}\,\ln\left(\frac{1}{4}\,\sqrt{\frac{\delta^2 + 12\,\delta_1}{3\,\lambda}}\,
\sinh\left(\frac{\sqrt{3\,\lambda}}{2}\,\tau\right)\right),\nonumber\\
\beta_{+}& = &\frac{2}{3}\,\frac{(\delta_2 - 2\,\delta_3)}{\sqrt{\nu^2 + 12\,\delta_1}}\,\ln\,
\tanh\left(\frac{\sqrt{3\,\lambda}}{2}\,\tau\right) + \frac{1}{6}\,\ln\,\left(\frac{1}{4}\,
\sqrt{\frac{\nu^2 + 12\,\delta_1}{3\,\lambda}}\right),\label{tolatetime-2}\\
\beta_{-}& = &-\frac{2}{\sqrt{3}}\,\frac{\delta_2}{\sqrt{\nu^2 + 12\,\delta_1}}\,\ln\,
\tanh\left(\frac{\sqrt{3\,\lambda}}{2}\,\tau\right) - \frac{1}{2\,\sqrt{3}}\,\ln\,\left(\frac{1}{4}\,
\sqrt{\frac{\nu^2 + 12\,\delta_1}{3\,\lambda}}\right).\nonumber
\end{eqnarray}
It is remarkable that  equations \eqref{solutions-fields} for the fields $\phi$ and $\sigma$ are maintained  for all
Bianchi Class A models, and in particular, when we use the gauge $N=e^{\beta_1+\beta_2+\beta_3}$, the solutions for these
fields are independent of the cosmological models, whose solutions are $\phi=\frac{\phi_0}{8} t$ and
$\sigma=\frac{\sigma_0}{8} t$.\\

After studying the $\gamma=0$ case we shall comment on the physical significance of these solutions.

\subsubsection{$\gamma=0$ case}
The Hamilton equations are
\begin{eqnarray}
\Pi_{1}^\prime=\Pi_{2}^\prime=\Pi_{3}^\prime&=& 4\Lambda e^{\beta_1+\beta_2+\beta_3}+\alpha_0 , \qquad \alpha_0=16\pi G\rho_0\nonumber\\
\Pi_\phi^\prime &=& 0, \qquad \Pi_\phi=\phi_1=cte, \label{pphi-0}\nonumber\\
%\label{sisgamce}\Pi_\sigma^\prime &=& 0, \qquad \Pi_\sigma=\sigma_1=cte,\nonumber\\
%\beta_1^\prime&=& \frac{1}{4}e^{-(\beta_1+\beta_2+\beta_3)}\left[-\Pi_1 +\Pi_2+\Pi_3 \right], \nonumber\\
%\beta_2^\prime&=& \frac{1}{4}e^{-(\beta_1+\beta_2+\beta_3)}\left[-\Pi_2 +\Pi_1+\Pi_3 \right], \\
%\beta_3^\prime&=& \frac{1}{4}e^{-(\beta_1+\beta_2+\beta_3)}\left[-\Pi_3 +\Pi_1+\Pi_2 \right], \nonumber\\
%\phi^\prime&=& \phi_2 \,e^{-(\beta_1+\beta_2+\beta_3)}, \nonumber\\
%\sigma^\prime&=& \sigma_2 \,e^{-(\beta_1+\beta_2+\beta_3)}.\nonumber
\end{eqnarray}

%To solve the case $\gamma=0$ we can see from the above system of equations \eqref{sisgamce} that
%\begin{equation}
%\Pi_1=\Pi_2 + \xi_2=\Pi_3 + \xi_3,
%\label{momentas-0}
%\end{equation}
%then re-introducing into the Hamiltonian equation \eqref{hami} we find one differential equation for the momenta $\Pi_1$ as
from which the momenta $\Pi_1$ fulfils the equation
\begin{equation}
\frac{1}{\Lambda}\Pi_1^{\prime 2} - 3\Pi_1^2 + \mu \Pi_1 + \xi_1=0,
\end{equation}
where
\begin{equation}
\xi_1=\delta_1 - \frac{\alpha^2_0}{\Lambda}.
\end{equation}
The  solution is given by
\begin{equation}
\Pi_{1}=\frac{\nu}{6} + \frac{1}{6}\,\sqrt{\nu^2 + 12\,\xi_{1}}\,\cosh\left(\sqrt{3\,\Lambda}\,\tau\right).
\end{equation}
Using this result  in the corresponding Hamilton equations, we obtain that
\begin{equation}
\beta_{1} + \beta_{2} + \beta_{3}= \ln\left(\frac{1}{8}\,\sqrt{\frac{\nu^2 + 12\,\xi_{1}}{3\Lambda}}\,
\sinh\left(\sqrt{3\,\Lambda}\,\tau\right)\right),
\end{equation}
and the corresponding solutions to the dilaton field $\phi$ and the moduli field $\sigma$ are given by
\begin{equation}
\phi=\frac{8\,\phi_{2}}{\sqrt{\nu^2 + 12\,\xi_{1}}}\,
\ln\tanh\left(\frac{\sqrt{3\Lambda}}{2}\tau\right),\,\,\,\,\,\,\,\,\,\,
\sigma=\frac{8\,\sigma_{2}}{\sqrt{\nu^2 + 12\,\xi_{1}}}\,
\ln\tanh\left(\frac{\sqrt{3\Lambda}}{2}\tau\right).
\label{soltomofi}
\end{equation}
So, the solutions to the Misner variables $(\Omega,\beta_\pm)$, are
\begin{eqnarray}
\Omega&=&\frac{1}{3}\,\ln\left(\frac{1}{8}\,\sqrt{\frac{\nu^2 + 12\,\xi_{1}}{3\Lambda}}\,\sinh\left(\sqrt{3\Lambda}\tau\right)\right),\nonumber\\
\beta_{+}&=&\frac{2}{3}\,\frac{\delta_{2} - 2\,\delta_{3}}{\sqrt{\nu^2 + 12\,\xi_{1}}}\,\ln\tanh\left(\frac{\sqrt{3\Lambda}}{2}\tau\right),\label{tolatetime-3}\\
\beta_{-}&=&-\frac{2}{\sqrt{3}}\,\frac{\delta_{2}}{\sqrt{\nu^2 +
12\,\xi_{1}}}\,\ln\tanh\left(\frac{\sqrt{3\Lambda}}{2}\tau\right).\nonumber
\end{eqnarray}

For both solutions, $\gamma=0, \pm 1$,  we also observe that for
positive (negative) $q_0$, the Universe expands slower (faster) than
in the absence of extra dimensions while the internal space shrinks
faster (slower) for positive (negative) $q_0$. In this case,
 isotropization parameters are also accordingly affected by $q_0$,
through relations \eqref{q0}.\\

%We can see that the solutions for the cases $\Lambda\not=0$, $\gamma=\pm 1$ and $\gamma=0$ are similar, only exist changes in
%the corresponding constants that we define in each cases, such as, $(\nu,\kappa,\mu)$ and $(_1, k_3 ,\xi_1)$. On the other
%hand, the volume associated to this model is grow up in all cases, it is possible view it from the expressions
%\eqref{tolatetime}, \eqref{tolatetime-1}, \eqref{tolatetime-2} and \eqref{tolatetime-3}, that will be relevant in the following
%analysis.\\

So far, our analysis has been completely classic. Now, in the next section we deal with the quantum scheme and we are going to
solve the WDW equation in standard quantum cosmology.\\

\section{Quantum scheme}

Solutions to the Wheeler-DeWitt (WDW) equation dealing with different problems have been extensively used in the literature. For example,
 the important quest of finding  a typical wave function for the Universe, was nicely addressed in \cite{Gibbons:1989ru}, while in
\cite{ASNA:ASNA2113110116} there appears an excellent summary concerning the problem of how a
Universe emerged from a big bang singularity,  cannot longer be neglected in the GUT epoch. On the other hand, the best candidates
for quantum solutions become those that have a damping behavior with respect to the scale factor, represented in our model with
the $\Omega$ parameter, in the sense that we obtain a good classical solution using the WKB approximation in any scenario in
the evolution of our Universe \cite{Hartle:1983ai}. The WDW equation for this model is achieved by replacing the momenta
$\Pi_{q^\mu}=-i\partial_{q^\mu}$, associated to the Misner variables $(\Omega,\beta_{+}, \beta_{-})$ and the moduli fields
$(\phi,\sigma)$ in the Hamiltonian \eqref{ham}. The factor $e^{-3\Omega}$ may be factor ordered with $\hat\Pi_\Omega$ in many
ways. Hartle and Hawking \cite{Hartle:1983ai} have suggested what might be called a semi-general factor ordering which in this
case would order $e^{-3\Omega}\hat\Pi^2_\Omega$ as
\begin{equation}
- e^{-(3- Q)\Omega}\, \partial_\Omega e^{-Q\Omega}
\partial_\Omega= - e^{-3\Omega}\, \partial^2_\Omega + Q\,
e^{-3\Omega} \partial_\Omega\,\,,
\label{fo}
\end{equation}
where  Q is any real constant that measure the ambiguity in the factor ordering in the variable $\Omega$ and the corresponding
momenta. We will assume in the following this factor ordering for the Wheeler-DeWitt equation, which becomes
\begin{equation}
\Box \, \Psi + Q \frac{\partial \Psi}{\partial
\Omega}-\frac{48}{11}\frac{\partial^2 \Psi}{\partial \phi^2}
-\frac{1}{11}\frac{\partial^2 \Psi}{\partial
\sigma^2}+\frac{18}{11}\frac{\partial^2 \Psi}{\partial\phi \partial
\sigma}  -\left[b_\gamma e^{3(1-\gamma)\Omega}+48\Lambda
e^{6\Omega}\right]\Psi =0, \label{WDW}
\end{equation}
where $\Box$ is the  three dimensional d'Lambertian in the $\ell^\mu=(\Omega,\beta_+,\beta_-)$ coordinates, with signature
$(- + +)$. On the other hand, we could interpreted the WDW equation \eqref{WDW} as a time-reparametrization invariance of the
wave function $\Psi$. At a glance, we can see that the WDW equation is static, this can be understood as the problem of time
in standard quantum cosmology. We can avoid this problem by measuring the physical time with respect a kind of time variable
anchored within the system, that means that we could understand the WDW equation as a correlation between the physical time and
a fictitious time \cite{Bae:2014aia,moniz2010quantum}. When we introduce the ansatz

\begin{equation}
\rm \Psi =\Phi(\phi,\sigma)\psi(\Omega,\beta_\pm),
\label{wavefunction}
\end{equation}
in \eqref{WDW}, we obtain the general set of differential equations
(under the assumed factor ordering),
\begin{eqnarray}
\Box \, \psi + Q \frac{\partial \psi}{\partial \Omega} -
\left[b_\gamma e^{3(1-\gamma)\Omega}+48\Lambda
e^{6\Omega}-\frac{\mu^2}{5}\right] \,
\psi &=&\rm 0,\label{wdwmod}\\
\frac{48}{11}\,\frac{\partial^2 \Phi}{\partial \phi^2}
+\frac{1}{11}\,\frac{\partial^2 \Phi}{\partial
\sigma^2}-\frac{18}{11}\,\frac{\partial^2 \Phi}{\partial\phi
\partial \sigma} +\mu^2 \Phi=0. \label {phi-1}
\end{eqnarray}
where we choose the separation constant $\frac{\mu^2}{5}$ for convenience to reduce the second equation. The solution to the
hyperbolic partial differential equation \eqref{phi-1}  is given by
\begin{equation}
\Phi(\phi,\sigma)= C_{1}\,\sin(C_{3}\,\phi + C_{4}\,\sigma + C_{5}) + C_{2}\,\cos(C_{3}\,\phi + C_{4}\,\sigma + C_{5}),
\label{solphi}
\end{equation}
where $\{C_i\}_{i=1}^5$ are integration constants and they are in terms of $\mu$. We claim that this solutions is the same for
all Bianchi Class A cosmological models, because the Hamiltonian operator in \eqref{WDW} can be written  in separated way as
$\hat H(\Omega,\beta_\pm,\phi,\sigma)\Psi=\hat H_g(\Omega,\beta_\pm)\Psi + \hat H_m(\phi,\sigma)\Psi=0$, where $\hat H_g$ y
$\hat H_m$ represents the Hamiltonian to gravitational sector and  the moduli fields, respectively. To solve equation
\eqref{wdwmod}, we now set $\psi(\Omega,\beta_{\pm})=\mathcal{A}(\Omega)\mathcal{B}_1(\beta_{+})\mathcal{B}_2(\beta_{-})$,
obtaining the following set of ordinary differential equations
\begin{align}
&\frac{d ^{2}{\cal A}}{d \Omega^2 }-Q\frac{d {\cal A}}{d \Omega }+\left[b_{\gamma} e^{-3(\gamma-1)\Omega}+48\Lambda e^{6\Omega}+\sigma^2\right]{\cal A}=0,\\
&\frac{d^{2}\mathcal{B}_{1}}{d\beta^{2}_{+}}+a_2^2\mathcal{B}_{1}=0, \qquad \Rightarrow \qquad {\cal B}_1=\eta_1 e^{i a_2 \beta_+}+\eta_2 e^{-i a_2 \beta_+},\\
&\frac{d^{2}\mathcal{B}_{2}}{d\beta^{2}_{-}}+a_3^2\mathcal{B}_{2}=0,\qquad
\Rightarrow \qquad {\cal B}_2=c_1 e^{i a_3\beta_-}+c_2e^{-i
a_3\beta_-},
\end{align}
where $\sigma^2=a_1^2+a_2^2+a_3^2$,  $c_i$ and $\eta_i$ are constants. We now focus on the $\Omega$ dependent part of the WDW
equation. We solve this equation when $\Lambda=0$ and $\Lambda\not=0$.
\begin{enumerate}
 \item{} For the case with null cosmological constant and $\gamma\neq 1$,
\begin{equation}
\frac{d^2\mathcal{A}}{d\Omega^2} - Q\,\frac{d\mathcal{A}}{d\Omega} + \left[b_{\gamma}\,e^{-3(1 - \gamma)\Omega}
+ \sigma^2\right]\mathcal{A}=0,
\label{aA-1}
\end{equation}
and by using the change of variable,
\begin{equation}
z=\frac{\sqrt{b_{\gamma}}}{p}\,e^{-\frac{3}{2}(\gamma - 1)\Omega},
\label{variable}
\end{equation}
where $p$ is a parameter to be determined, we have
\begin{eqnarray*}
\frac{d\mathcal{A}}{d\Omega}&=&\frac{d\mathcal{A}}{dz}\frac{dz}{d\Omega}=-\frac{3}{2}\,\left(\gamma - 1\right)
\,z\,\frac{d\mathcal{A}}{dz},\\
\frac{d^2\mathcal{A}}{d\Omega^2}&=&\frac{9}{4}\,\left(\gamma - 1\right)^2\,z^2\,\frac{d^2\mathcal{A}}{dz^2} +
\frac{9}{4}\,\left(\gamma - 1\right)^2\,z\,\frac{d\mathcal{A}}{dz}.
\end{eqnarray*}
Hence, we arrive at the equation
\begin{equation}
\frac{9}{4}\left(\gamma - 1\right)^2\,z^2\,\frac{d^2\mathcal{A}}{dz^2} + \frac{9}{4}\left(\gamma - 1\right)^2\,z\,\frac{d\mathcal{A}}{dz}
- \frac{3}{2}\,Q\,\left(\gamma - 1\right)\,z\,\frac{d\mathcal{A}}{dz} + \left[p^2\,z^2 + \sigma^2\right]=0,
\label{t1}
\end{equation}
where we have assumed that $\mathcal{A}$ is of the form \cite{zaitsev2002handbook}
\begin{equation}
\mathcal{A}=z^{q\,Q}\,\Phi(z),
\label{transformationa}
\end{equation}
with $q$ is yet to be determined. We thus get after substituting in \eqref{t1},
\begin{equation*}
\frac{9}{4}\biggl(\gamma - 1\biggr)^2\,z^{q\,Q}\biggl[z^2\,\frac{d^2\Phi}{dz^2} + z\biggl(1 + 2\,q\,Q + \frac{2}{3}
\frac{Q}{\gamma -1}\biggr)\frac{d\Phi}{dz} + \biggl(\frac{4p^2}{9(\gamma-1)^2}\,z^2
+ Q^2\biggl\{q^2 +\frac{2}{3}\frac{q}{\gamma -1}\biggr\}
+ \frac{4\sigma^2}{9(\gamma -1)^2}\biggr)\Phi\biggr]=0,
\end{equation*}
which can be written as
\begin{equation}
z^2\,\frac{d^2\Phi}{dz^2} + z\,\frac{d\Phi}{dz} + \left[z^2 - \frac{1}{9(\gamma - 1)^2}\left(Q^2 - 4\sigma^2\right)\right]\Phi=0,
\end{equation}
which is the Bessel differential equation for the function $\Phi$ when $p$ and $q$ are fixed to
\begin{equation}
q=-\frac{1}{3(\gamma - 1)},\qquad p=\frac{3}{2}|\gamma-1|,
\end{equation}
which in turn means that the transformations \eqref{variable} and \eqref{transformationa} are
\begin{equation*}
z=\frac{2\sqrt{b_{\gamma}}}{3|\gamma - 1|}\,e^{-\frac{3}{2}(\gamma - 1)\Omega},\qquad
\mathcal{A}=z^{-\frac{Q}{3(\gamma - 1)}\Phi(z)}.
\end{equation*}
Hence, the solution for \eqref{aA-1} is of the form
\begin{equation}
\mathcal{A}_{\gamma}=c_{\gamma}\left(\frac{2\sqrt{b_{\gamma}}}{3|\gamma
- 1|}\,e^{-\frac{3}{2}(\gamma - 1)\Omega}\right)
^{-\frac{Q}{3(\gamma -
1)}}Z_{i\nu}\left(\frac{2\sqrt{b_{\gamma}}}{3|\gamma -
1|}\,e^{-\frac{3}{2}(\gamma - 1)\Omega}\right),
\end{equation}
with
\begin{equation}
\nu=\pm\sqrt{\frac{1}{9(\gamma - 1)^2}( 4\sigma^2-Q^2)},
\end{equation}
where $Z_{i\nu}=J_{i\nu}$ is the ordinary Bessel function with imaginary order. For the particular case when the factor
ordering is zero, we can easily construct a wave packet \cite{Hartle:1983ai,Kiefer:1989va,Kiefer:1988tr} using the identity
\begin{equation*}
\int_{-\infty}^\infty \text{sech}\left(\frac{\pi\eta}{2}\right)\,
J_{i\eta}(z)\,d\eta=2\,\sin(z),
\end{equation*}
so, the total wave function becomes
\begin{equation*}
\Psi(\Omega,\beta_\pm,\phi,\sigma)=\Phi(\phi,\sigma)\,
\sin\left(\frac{2\sqrt{b_{\gamma}}}{3|\gamma -
1|}\,e^{-\frac{3}{2}(\gamma - 1)\Omega}\right)\left[\eta_1 e^{i a_2
\beta_+}+\eta_2 e^{-i a_2 \beta_+}\right] \left[c_1 e^{i
a_3\beta_-}+c_2e^{-i a_3\beta_-} \right]
\end{equation*}
Notice that the influence of extra dimensions through the presence of the moduli $\phi$ and $\sigma$ appears in the solution in the amplitud of the wave function.
 We shall comment on this later on.

\item{} Case with null cosmological constant and $\gamma=1$.
For this case we have

\begin{equation}
\frac{d^2\mathcal{A}_{1}}{d\Omega^2} - Q\,\frac{d\mathcal{A}_{1}}{d\Omega} + \sigma_{1}^2\,\mathcal{A}_{1}=0, \qquad
\sigma^2_{1}=b_{1} + \sigma^2,
\label{aA-2}
\end{equation}
whose solution is
\begin{equation}
\mathcal{A}_{1}=A_{1}\,e^{\frac{Q + \sqrt{Q^2 + 4\sigma^2_{1}}}{2}\Omega} +
A_{2}\,e^{\frac{Q - \sqrt{Q^2 + 4\sigma^2_{1}}}{2}\Omega}.
\end{equation}
so, the total wave function becomes
\begin{equation*}
\Psi(\Omega,\beta_\pm,\phi,\sigma)=\Phi(\phi,\sigma)\, \left[A_1
e^{\frac{Q + \sqrt{Q^2 + 4\sigma^2_{1}}}{2}\Omega} +
A_{2}\,e^{\frac{Q - \sqrt{Q^2 +
4\sigma^2_{1}}}{2}\Omega}\right]\left[\eta_1 e^{i a_2
\beta_+}+\eta_2 e^{-i a_2 \beta_+}\right] \left[c_1 e^{i
a_3\beta_-}+c_2e^{-i a_3\beta_-} \right].
\end{equation*}

\item{} If we include the cosmological constant term for this particular case, we have (Zaitsev 2002)

\begin{equation}
\mathcal{A}=e^{\frac{Q}{2}\,\Omega}\,Z_{\nu}\left(4\sqrt{\frac{\Lambda}{3}}\,e^{3\Omega}\right), \qquad
\nu=\pm\frac{1}{6}\sqrt{Q^2 + 4\sigma^2_{1}},
\end{equation}
where $\Lambda>0$ in order to have ordinary Bessel functions as solutions, in other case, we will have the modified Bessel
function. When the factor ordering parameter Q equals zero, we have the same generic Bessel functions as solutions, having the
imaginary order $\nu=\pm i\frac{\sqrt{\sigma_1^2}}{3}$.

\item{} $\gamma=-1$ and $\Lambda \neq0$ and any factor ordering Q.

\begin{equation}
\frac{d^2\mathcal{A}_{-1}}{d\Omega^2} - Q\frac{d\mathcal{A}_{-1}}{d\Omega}
+ \left[d_{-1} e^{6\Omega} - \sigma^2\right]\mathcal{A}_{-1}=0,
\qquad d_{-1}=48\mu_{-1} + 48\Lambda
\label{A-1}
\end{equation}
with the solution
\begin{equation}
\mathcal{A}_{-1}=\left(\frac{\sqrt{d_{-1}}}{3}e^{3\Omega}\right)^{\frac{Q}{6}}
Z_{\nu}\left(\frac{\sqrt{d_{-1}}}{3}e^{3\Omega}\right), \qquad
\nu=\pm\frac{1}{6}\sqrt{Q^2 + 4\sigma^2}
\end{equation}
where $Z_\nu$ is a generic Bessel function. When $b_{-1}>0$, we have ordinary Bessel functions, in other case we have modified
Bessel functions.

\item{} $\gamma=0$, $\Lambda \neq0$ and factor ordering  $Q=0$

\begin{equation}
\frac{d^2\mathcal{A}_0}{d\Omega^2 } + \left(48\Lambda e^{6\Omega} + b_0 e^{3\Omega} + \sigma^2\right)\mathcal{A}_0=0,
\qquad b_{0}=48\mu_{0}
\label{C0}
\end{equation}
with the solution
\begin{equation}
\mathcal{A}_0 = e^{-3\Omega/2}\left[D_{1}M_{-\frac{i b_0}{24\sqrt{3\Lambda}},
-\frac{i\sigma}{3}}\left(\frac{8i\,e^{3\Omega}\sqrt{\Lambda}}{\sqrt{3}}\right)
+ D_{2}W_{-\frac{i b_0}{24\sqrt{3\Lambda}},-\frac{i\sigma}{3}}\left(\frac{8ie^{3\Omega}\sqrt{\Lambda}}{\sqrt{3}}\right)\right],
\end{equation}
where $M_{k,p}$ And $W_{k,p}$ are Wittaker functions and $D_i$ are integration constants.

\end{enumerate}

\section{Final Remarks}
In this work we have explored a compactification of a
ten-dimensional gravity theory coupled with a time-dependent dilaton
into a time-dependent six-dimensional torus. The effective theory
which emerges though this process resembles in the Einstein frame to
that described by the Bianchi I model. By incorporating the
barotropic matter and cosmological content and by using the
analytical procedure of Hamilton equation of classical mechanics, in
appropriate coordinates, we found the classical solution for the
anisotropic Bianchi type I cosmological models. In particular, the
Bianchi type I is complete solved without using a particular gauge.
With these solutions we can validate our qualitative analysis on
isotropization of the cosmological model, implying that this become
when the volume
is large in the corresponding \emph{time} evolution. \\

We find that for all  cases, the Universe expansion
is gathered as time runs while the internal space shrinks.
Qualitatively this models shows us that extra dimensions are forced
to decrease its volume for an expanding Universe. Also we notice
that the presence of extra dimensions affects how fast the Universe (with matter)
expands through the presence of a constant related to the moduli
momenta. This is not unexpected since we are not considering a
potential for the moduli, which implies that they are not stabilized
and consequently an effective model should only take into account
their constant momenta. We observe that isotropization is not
affected in the cases without matter, but in the matter presence,
isotropization can be favored or retarded according to how fast the
moduli evolve with respect to each other.\\

Could be interesting to study other type of matter in this context, beyond the barotropic matter. For
instance, the Chaplygin gas, where  a proper time,  characteristic
to this matter, leads to the presence of singularities type I, II, III and IV (generalizations to these models are presented in
\cite{bamba,nojiri,jambrina}) which also appear
in the phantom scenario to dark/energy matter. In our case, since the matter we are considering is barotropic, initial singularities of these type  do not emerge from our analysis, implying also that
phantom fields are absent. It is important to remark, that this model is a very simplified one in the sense that we do not consider the presence of moduli dependent matter and we do not analyse under which conditions inflation is present or how it starts. We plan to study this important task in a different work.
 \\

Concerning  the quantum
scheme we can observe that this anisotropic model is completely
integrable without employ numerical methods, similar solutions to
 partial differential equation into the  gravitational variables has been
found in \cite{SOCORRO2010}, and we obtain that the solutions in the
moduli fields are the same for all Bianchi Class A cosmological
models, because the Hamiltonian operator in \eqref{WDW} can be
written  in separated way as $\hat
H(\Omega,\beta_\pm,\phi,\sigma)\Psi=\hat H_g(\Omega,\beta_\pm)\Psi +
\hat H_m(\phi,\sigma)\Psi=0$, where $\hat H_g$ y $\hat H_m$
represents the Hamiltonian to gravitational sector and  the moduli
fields, respectively, with the full wave function given by $\rm \Psi
=\Phi(\phi,\sigma) \Theta(\Omega,\beta_\pm)$. In order to have the
best candidates for quantum solutions become those that have a
damping behavior with respect to the scale factor, represented in
our model with the $\Omega$ parameter, in this way we shall dropped
in the full solution the modified Bessel function $\rm I_\nu(z)$ or
Bessel function $\rm Y_\nu(z)$, with $\nu$ the order of these
functions.\\

We can observe that the quantum solution in the  $\Omega$ sector is
similar to the corresponding  FRW cosmological
 model, found in different schemes \cite{Cavaglia:2000rx, Cavaglia:2000uu,MartinezPrieto:2004fd,Socorro:2003fn,moniz2010quantum}.
 Also, similar analysis can be found in recent published paper by two authors of this work \cite{JL}.\\

The presence of extra dimensions could have a stronger influence on
the isotropization process by having a moduli-dependent potential,
which can be gathered by turning on extra fields in the internal
space, called fluxes. It will be interesting to consider a more
complete compactification process in which all moduli are considered
as time-dependent fields as well as time-dependent fluxes. We leave
these important analysis for future work.

\acknowledgments{ \noindent This work was partially supported by
CONACYT  167335, 179881 grants. PROMEP grants UGTO-CA-3 and DAIP-UG 640/2015. This work
is part of the collaboration within the Instituto Avanzado de
Cosmolog\'{\i}a and Red PROMEP: Gravitation and Mathematical Physics
under project \emph{Quantum aspects of gravity in cosmological
models, phenomenology and geometry of space-time}. One of authors
(LTS) was supported by a PhD scholarship in the graduate program by
CONACYT.

\appendix
\section{Dimensional reduction} \label{appex}
With the purpose to be consistent, here we  present the main ideas to  dimensional
reduce  the action \eqref{action}.  By the use of the conformal transformation
\begin{equation}
\hat{G}_{MN}=e^{\Phi/2}\,G^{E}_{MN},
\label{weyltrans}
\end{equation}
the action \eqref{action} can be written as
\begin{equation}
S=\frac{1}{2\kappa^2_{10}}\,\int\,d^{10}X\,\sqrt{-\hat{G}}\,\biggl(e^{\Phi/2}\,\mathcal{\hat{R}}
+ 4\,G^{E_{MN}}\,\nabla_{M}\Phi\nabla_{N}\Phi\biggl) + \int\,d^{10}X\,\sqrt{-G^{E}}\,e^{5\Phi/2}\,\mathcal{L}_{\text{matt}},
\label{sopas}
\end{equation}
where the ten-dimensional scalar curvature $\mathcal{\hat{R}}$ transform according to the conformal transformation as
\begin{equation}
\mathcal{\hat{R}}=e^{-\Phi/2}\biggl(\mathcal{R}^{E} - \frac{9}{2}\,G^{E_{MN}}\,\nabla_{M}\nabla_{N}\Phi - \frac{9}{2}\,G^{E_{MN}}
\,\nabla_{M}\Phi\nabla_{N}\Phi\biggr).
\label{sopita}
\end{equation}
By substituting the expression \eqref{sopita} in \eqref{sopas} we obtain
\begin{equation}
S=\frac{1}{2\kappa^2_{10}}\int d^{10}X\,\sqrt{-G^{E}}\,\biggl(\mathcal{R}^{E} - \frac{9}{2}\,G^{E_{MN}}\,\nabla_{M}\nabla_{N}\Phi
- \frac{1}{2}\,G^{E_{MN}}\,\nabla_{M}\Phi\nabla_{N}\Phi\biggr) + \int d^{10}X\,\sqrt{-G^{E}}\,e^{5\Phi/2}\,\mathcal{L}_{\text{matt}}.
\label{actein}
\end{equation}

The last expression is the ten-dimensional action in the Einstein
frame.  Expressing the metric determinant in four-dimensions in terms of the moduli field $\sigma$ we have
\begin{equation}
\text{det}\,\hat{G}_{MN}=\hat{G}=e^{-12\sigma}\,\hat{g}.
\label{dete}
\end{equation}

By substituting the last expression \eqref{dete} in \eqref{action} and considering that
\begin{align}
\mathcal{\hat{R}}^{(10)}=\hat{\mathcal{R}}^{(4)} - 42\,\hat{g}^{\mu\nu}\,\nabla_{\mu}\sigma\nabla_{\nu}\sigma
+ 12\,\hat{g}^{\mu\nu}\,\nabla_{\mu}\nabla_{\nu}\sigma,
\label{zehnricci}
\end{align}
we obtain that
\begin{align}
S=\frac{1}{2\kappa^2_{10}}\int d^4x\,d^6y\,\sqrt{-\hat{g}}\,e^{-6\sigma}e^{-2\Phi}\bigg[\hat{\mathcal{R}}^{(4)}
- 42\,\hat{g}^{\mu\nu}\,\nabla_{\mu}\sigma\nabla_{\nu}\sigma + 12\,\hat{g}^{\mu\nu}\,\nabla_{\mu}\nabla_{\nu}\sigma
+ 4\,\hat{g}^{\mu\nu}\,\nabla_{\mu}\Phi\nabla_{\nu}\Phi\biggr].
\label{prelim}
\end{align}
Let us redefine in the last expression the dilaton field $\Phi$ as
\begin{equation}
\Phi= \phi - \frac{1}{2}\,\ln(\hat{V}),
\label{volu}
\end{equation}
where $\hat{V}=\int d^6y$. So, the expression \eqref{prelim} can be
written as
\begin{equation}
S=\frac{1}{2\kappa^2_{10}}\int d^4x\,\sqrt{-\hat{g}}\,e^{-2(\phi + 3\sigma)}\bigg[\hat{\mathcal{R}}^{(4)}
- 42\,\hat{g}^{\mu\nu}\,\nabla_{\mu}\sigma\nabla_{\nu}\sigma + 12\,\hat{g}^{\mu\nu}\,\nabla_{\mu}\nabla_{\nu}\sigma
+ 4\,\hat{g}^{\mu\nu}\,\nabla_{\mu}\phi\nabla_{\nu}\phi\biggr].
\label{prelim1}
\end{equation}
The four-dimensional metric $\hat{g}_{\mu\nu}$ means that the metric in the String frame, we should take a conformal transformation
linking the String and Einstein frames. We label by $g_{\mu\nu}$ the metric of the external space in the Einstein frame, this
conformal transformation is given by
\begin{equation}
\hat{g}_{\mu\nu}=e^{2\Theta}\,g_{\mu\nu},
\label{trans}
\end{equation}
which after some  algebra gives the four-dimensional scalar curvature
\begin{equation}
\hat{\mathcal{R}}^{(4)}=e^{-2\Theta}\biggl(\mathcal{R}^{(4)} - 6\,g^{\mu\nu}\,\nabla_{\mu}\nabla_{\nu}\Theta
- 6\,g^{\mu\nu}\,\nabla_{\mu}\Theta\nabla_{\nu}\Theta\biggr),
\label{vierscalr}
\end{equation}
where the function $\Theta$ is given by the transformation
\begin{equation}
\Theta = \phi + 3\sigma + \ln\left(\frac{\kappa^2_{10}}{\kappa^2_{4}}\right).
\label{transtotal}
\end{equation}
Now, we must replace the expressions \eqref{trans}, \eqref{vierscalr}, \eqref{transtotal} in the expression \eqref{prelim1} we
find
\begin{align}
S=\frac{1}{2\kappa^2_4}\int d^4x\,\sqrt{-g}\,\biggl(\mathcal{R} -6\,g^{\mu\nu}\,\nabla_{\mu}\nabla_{\nu}\phi
-6\,g^{\mu\nu}\,\nabla_{\mu}\nabla_{\nu}\sigma - 2\,g^{\mu\nu}\,\nabla_{\mu}\phi\nabla_{\nu}\phi - 96\,g^{\mu\nu}\,\nabla_{\mu}\sigma
\nabla_{\nu}\sigma - 36\,g^{\mu\nu}\,\nabla_{\mu}\phi\nabla_{\nu}\sigma\biggl).
\label{deduction}
\end{align}
At first glance, is important to clarify one point related with the stress-energy tensor which has the matrix form
\eqref{matrixform}; this tensor belongs to the String frame. In order to write the stress-energy tensor in the Einstein frame
we need to work with the last integrand of the expression \eqref{action}. So, after taking the variation with respect
to the ten-dimensional metric and opening the expression we see that
\begin{align*}
\int\,d^{10}X\,\sqrt{-\hat{G}}\,e^{-2\Phi}\,\kappa^2_{10}\,e^{2\Phi}\,\hat{T}_{MN}\,
\hat{G}^{MN}=
\int\,d^{4}x\,\sqrt{-g}\,e^{2(\Theta+\phi)}\,\hat{T}_{MN}\,\hat{G}^{MN},\nonumber\\
=\int\,d^{4}x\,\sqrt{-g}\,\biggl(e^{2\phi}\,\hat{T}_{\mu\nu}\,
g^{\mu\nu} + e^{2(\Theta + \phi)}\,\hat{T}_{mn}\,
\hat{g}^{mn}\biggr),
\end{align*}
where we can observe that that the four-dimensional stress-energy tensor in the Einstein frame is defined as we said in the expression
\eqref{hauptarroz}.

\end{document}